\documentclass[conference]{IEEEtran}

\makeatletter
\def\ps@IEEEtitlepagestyle{%
  \def\@oddfoot{\mycopyrightnotice}%
  \def\@evenfoot{}%
}
\def\mycopyrightnotice{%
  {\footnotesize 978-1-6654-7095-7/22/\$31.00~\copyright~2022 IEEE\hfill}
  \gdef\mycopyrightnotice{}
}

\usepackage{blindtext}
\usepackage{eso-pic}
\IEEEoverridecommandlockouts
\usepackage{cite}
\usepackage{amsmath,amssymb,amsfonts}
\usepackage{algorithmic}
\usepackage{graphicx}
\usepackage{textcomp}
\usepackage{xcolor}
\def\BibTeX{{\rm B\kern-.05em{\sc i\kern-.025em b}\kern-.08em
    T\kern-.1667em\lower.7ex\hbox{E}\kern-.125emX}}
    
\usepackage{eso-pic}
\newcommand\AtPageUpperMyright[1]{\AtPageUpperLeft{%
 \put(\LenToUnit{0.17\paperwidth},\LenToUnit{-2cm}){%
     \parbox{0.9\textwidth}{\raggedleft\fontsize{8}{11}\selectfont #1}}%
 }}%
\newcommand{\conf}[1]{%
\AddToShipoutPictureBG*{%
\AtPageUpperMyright{#1}
}
}    
    
\usepackage{tikz}
\usepackage{pgf}
\usepackage{pgfplots}
\usepackage{multirow} 
\usepackage{listings} 
\usepackage{subcaption} 
\usetikzlibrary{calc}   

\newcommand\bstrut{\rule[-1.0ex]{0pt}{0pt}}

\begin{document}
\title{\vspace*{1cm} Road Network Variation Based on HD Map Analysis for the Simulative Safety Assurance of Automated Vehicles\\
\thanks{This research is funded by the VVM project research initiative, promoted by the German Federal Ministry for Economic Affairs and Climate Action (BMWK).}
}

\author{\IEEEauthorblockN{1\textsuperscript{st} Daniel Becker}
\IEEEauthorblockA{\textit{Institute for Automotive Engineering} \\
\textit{RWTH Aachen University}\\
Aachen, Germany \\
daniel.becker@ika.rwth-aachen.de}
\and
\IEEEauthorblockN{2\textsuperscript{nd} Christian Geller}
\IEEEauthorblockA{\textit{Institute for Automotive Engineering} \\
\textit{RWTH Aachen University}\\
Aachen, Germany \\
christian.geller@ika.rwth-aachen.de}
\and
\IEEEauthorblockN{3\textsuperscript{rd} Lutz Eckstein}
\IEEEauthorblockA{\textit{Institute for Automotive Engineering} \\
\textit{RWTH Aachen University}\\
Aachen, Germany \\
lutz.eckstein@ika.rwth-aachen.de}
}

\maketitle
\conf{\textit{  Proc. of the International Conference on Electrical, Computer, Communications and Mechatronics Engineering  (ICECCME) \\ 
16-18 November 2022, Maldives}}
\begin{abstract}
The validation and verification of automated driving functions (ADFs) is a challenging task on the journey of making those functions available to the public beyond the current research context. Simulation is a valuable building block for scenario-based testing that can help to model traffic situations that are relevant for ADFs. In addition to the surrounding traffic and environment of the ADF under test, the logical description and automated generation of concrete road networks have an important role. We aim to reduce efforts for manual map generation and to improve the automated testing process during development.

Hence, this paper proposes a method to analyze real road networks and extract relevant parameters for the variation of synthetic simulation maps that correspond to real-world properties. Consequently, characteristics for inner-city junctions are selected from Here HD map. Then, parameter distributions are determined, analyzed and used to generate variations of road networks in the OpenDRIVE standard. The presented methodology enables efficient road network modeling which can be used for large scale simulations. The developed road network generation tool is publicly available on GitHub.
  
\end{abstract}


\section{Introduction}\label{sec_intro}
The safety assurance for automated driving functions (ADFs) and advanced driver assistant systems (ADAS) poses a major challenge when releasing those functions to the public \cite{stellet2019validation}. Research shows that the classical validation methods of automotive functions, such as test drives on public roads, would require driving a considerable amount of test kilometers which would be economically and in terms of time not acceptable to be executed \cite{wachenfeld2016release}. Therefore, the concept of scenario-based testing has been proposed e.g., in \cite{schuldt2017beitrag}. This approach has been further investigated in various research projects such as SET Level \cite{setlevel}, and VVMethoden \cite{vvm}.

Simulation approaches are an increasingly valuable building block in the overall verification and validation process of an ADF. In contrast to real-world test drives, this method saves costs and shortens the test time since simulations can be executed faster than in real time. In addition, detecting functional errors at an early development stage is possible, since a reproducible simulation test case is designed to cover predefined functional requirements. It is also appropriate to create any number of scenarios and explore critical driving situations without endangering people or risking the loss of test vehicles \cite{HanduchFASsim}. 

\begin{figure}[t] 		
	\centering
	\includegraphics{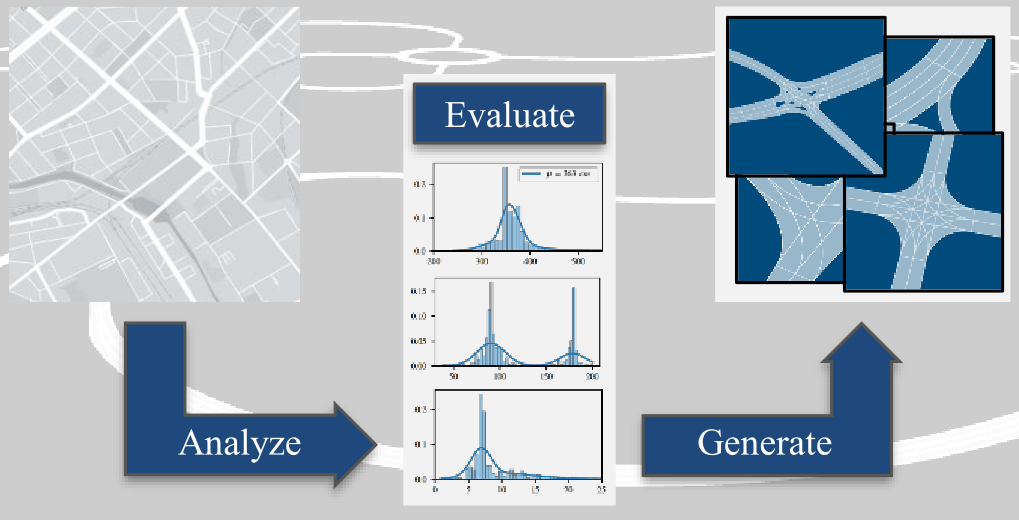}
	\caption{Overview of the workflow for the generation of synthetic road networks. Data from real roads are analyzed, and parameters including their probability distributions are extracted. Then, a set of road networks corresponding to the same logical template is generated.}
	\label{fig_motivation}
\end{figure}

In the context of scenario-based testing within various simulation tools, it is necessary to establish a common understanding of the term scenario to make them executable on different platforms. One approach to structure the setup of a scenario is the so-called six-layer model, which has been published in \cite{schuldt2017beitrag} and then has been further developed by \cite{bock2018data} and \cite{scholtes20206}. As described in \cite{becker2020roadgen}, besides the moving objects of a scenario, the static part (i.e. the road network) of a scenario should be variable and logically described as well. Since simulated road networks are solely an abstraction of the real world, the possibility arises to create synthetic junctions and roads instead of trying to replicate actual existing road networks. Which is considerably less effort than modeling real world junctions and can be done automatically. However, these generated simulation maps should somehow be as realistic as possible.

Hence, this paper presents an approach to analyze real-world HD maps, extract relevant parameter distributions, and finally generate variations of road networks for the safety assurance process of an ADF. An overview of the overall workflow is shown in Fig.~\ref{fig_motivation}.
The paper is structured as follows: Section~\ref{sec_relatedWork} gives a brief overview of simulation maps and road generation. In Section~\ref{sec_method}, the overall methodology that has been pursued is described. The data extraction and analysis of junctions are discussed in Section~\ref{sec_dataextraction}, and Section~\ref{sec_eval} deals with the data extraction results and shows possible inputs for the road generation. Section~\ref{sec_roadgen} describes the necessary extensions to the existing road generation methodology. Finally, Section~\ref{sec_conclusion} gives a conclusion on the presented work and an outlook on possible future work. 

Our main contributions are as follows:
\begin{itemize}
 \item We propose a methodology for parameter extraction from HD maps to extract characteristic intersection properties.
 \item We further develop our road generation method to enable the creation of stochastic intersections in OpenDRIVE.
 \item For evaluation, we apply the approach to German cities, analyze the intersection parameters, and generate road networks by parameter sampling. 
\end{itemize}

\section{Related work}\label{sec_relatedWork}
As introduced before, the methodology of this work enables road network variations based on real map data. Due to different stages in the development process, we also conduct related work in separate steps. Besides an overview for scenario description, we mainly investigate the parameter extraction from map data, as well as the route generation itself.

\subsection{Standardization Formats}\label{sec_relatedWorl_scenario}
In recent years, scenario-based testing has become popular for validating and verifying automated vehicles. However, to make scenarios interchangeable among different simulation platforms, efforts have been made to standardize the description of scenarios by the organization ASAM e.V. The road network and some parts of layers two and three may be described in the OpenDRIVE \cite{Opendrive} format. The remaining top layers are described in the OpenSCENARIO \cite{osc10} format. These activities together have the working title ASAM OpenX~\cite{asamOpenX}.

\subsection{Map Data Extraction}\label{sec_relatedWorl_map}
To create realistic road layouts, real-world roads can to be analyzed based on digital maps. Possible data sources are OpenStreetMap (OSM)~\footnote{https://www.openstreetmap.org}, Google Maps~\footnote{https://www.google.de/maps}, or Here HD map~\footnote{https://www.here.com/platform/automotive-services/hd-maps}, each with a different level of detail, accuracy, and coverage. Road data can be provided and maintained by the community, authorities, or companies.

Several approaches have been developed to extract relevant parameters from HD maps. E.g., \cite{hillerRoadAggr} presents a data extraction and analysis based on OSM data. The authors analyze motorway metadata such as the number of lanes or speed limits, compare them to the regulations that exist for German motorways, and measure the coverage in datasets of those metadata.

\subsection{Road Generation}\label{sec_relatedWorl_generation}
In general, several approaches exist for the automatic generation of road networks, each with a different focus. On the one hand, methodologies enable the exact modeling of real-world road networks as exact as possible. This approach automatically introduces variety, which is a crucial effect during testing. Well-known techniques are \cite{Despine11}, or \cite{Cura15}, capable of generating real-world intersections but suffer when replicating exact lane structures. Another issue for real-world data approaches is the lack of available data and, therefore, the resulting manual validation effort.
On the other hand, road generation approaches such as PGDrive\cite{Li20} or Pyodrx\cite{Pyodrx} model artificial intersections based on predefined parameters. Even more advanced approaches are JunctionArt \cite{muktadirrealistic} or \cite{becker2020roadgen}. In all approaches, the authors introduced a concept for the logical description of road networks, which allows a simple and parametric generation of simulation maps in the OpenDRIVE \cite{Opendrive} format.


\section{Methodology}\label{sec_method}
This paper proposes an approach to generate randomly varied synthetic road networks. To find meaningful values for the parameterization of such a road network, real-world data can be analyzed. As already described in \ref{sec_relatedWorl_map}, different data sources and map formats are available. We used Here HD map as data source since in contrast to OSM or Google Maps the available precision on lane geometries is at centimeter-level.

As a next step, the extracted junction data are evaluated regarding their probability distributions, followed by an analysis of which parameters are suitable for the generation of synthetic road networks. In addition, federal road-building regulations \cite{baier2007richtlinien} can be considered to validate that the extracted parameter distributions (e.g. for the lane width inside a junction) are reasonable.

As a final step, the methodology is applied to an existing road generation tool (cf.~\cite{becker2020roadgen}) which is extended by a functionality that specifies distinct parameters as random variables. Further, those variables can be linear combined to be able to depict possible dependencies of existing road networks. 

\section{Data Extraction and Analysis}\label{sec_dataextraction}
This section describes the extraction of data from Here HD map. First, the input format is briefly introduced. Then, the possible parameters that can be extracted are outlined.

\subsection{Here HD Map}\label{sec_heremaps}
\begin{figure}
	\begin{subfigure}{0.49\columnwidth}	
		\centering
		\includegraphics[width=\columnwidth]{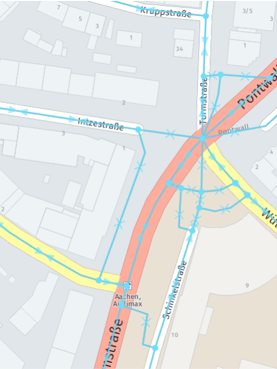}
		\subcaption{Topology geometry layer}
	\end{subfigure}
	\begin{subfigure}{0.49\columnwidth}	
		\centering
		\includegraphics[width=\columnwidth]{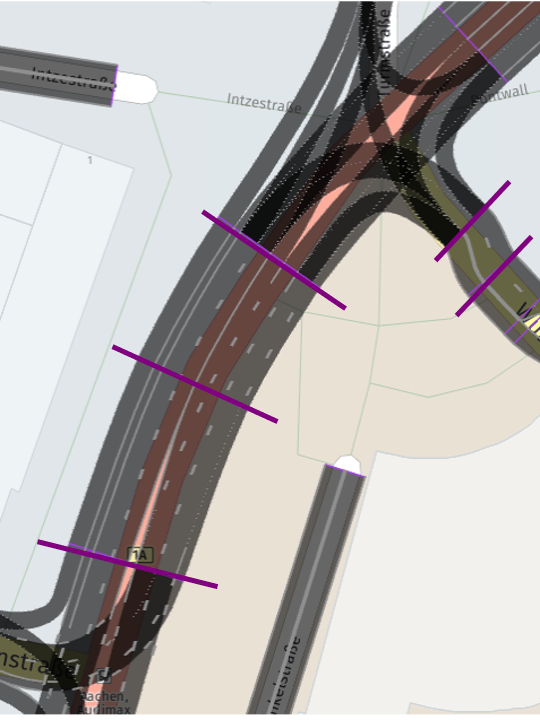}
		\subcaption{Lane geometry polyline layer}
	\end{subfigure}
	\caption{Two layers of Here HD map that help analyzing a road network. The topology geometry (a) consists of nodes and links that describe the road's logical course. The lane geometry polyline layer (b) describes the geometry of each lane along the road. Lane geometries are provided with a higher resolution than the topology information. \copyright www.here.com}
	\label{fig_here_layers}
\end{figure}

The basis of the analysis in this work is data from Here HD map~\cite{heremaps} which is organized in a layer structure. The macroscopic course of the road network is described in the \textit{Road Centerline Model} within the \textit{topology-geometry} layer (cf. Fig.~\ref{fig_here_layers} (a)) which is comparable to data in OpenStreepMap. It consists of nodes and links between them. The layer can be used to get an understanding of how to navigate through a road network and does not consist of high precision coordinates. To get the latter, Here HD map provides the \textit{HD Lane Model} in the \textit{lane-geometry-polyline} layer (cf. Fig.~\ref{fig_here_layers} (b)) which is lying on top of the topology layer. In addition to those two layers, for this study, the following layers were extracted and used to calculate intersection parameter distributions:  
\begin{itemize}
	\item \textbf{routing-attributes}: Links attributes such as the authorized vehicles, the presence of e.g. traffic lights, or buildings alongside the road.
	\item \textbf{topology-geometry}: List of links and nodes, their geometry, and the way they are connected to each other.
	\item \textbf{speed-attributes}: Speed limit along the link.
	\item \textbf{lane-geometry-polyline}: High-resolution geometry of each lane encapsulated as lane groups.
	\item \textbf{lane-attributes}: Attributes of each lane.
	\item \textbf{lane-topology}: Topological information on how lane groups are connected to each other at the junction and between two lane groups.
	\item \textbf{lane-road-references}: Description of how lane groups and road topology are alaid over one another.
\end{itemize}

Further, the map is spatially divided into rectangular tiles of approximately $1.5$km x $2.5$km (can vary depending on latitude/longitude). The analysis is done tile by tile and is described in the following section.

\subsection{Intersection Data Analysis}\label{sec_data_analysis}
The data processing is split into two separate modules: Intersection extraction and intersection analysis. The first module creates a list of all the intersections within given tiles. In addition, further parameters such as distance between junctions are computed. The second module then merges multiple processed tiles, categorizes the intersections, and generates new parameters to describe the junction more precisely, e.g. the maximum curvature of turning lanes. In the following, both modules are briefly described.

\subsubsection{Intersection Extraction}
This module processes a list of tiles and extracts intersections based on the \textit{Road Centerline Model}. All links in the topology-geometry-layer are processed, and it is decided whether they belong to an intersection or not. The intermediate result is a list of intersections with corresponding nodes from the \textit{Road Centerline Model}.

Further, the relation to the lane groups of the \textit{HD Lane Model} is calculated such that specific lane courses can be analyzed in later steps. Finally, all identified intersections are equipped with metadata from the layers described in Section~\ref{sec_heremaps} such as the number of incoming lanes per road or the presence of traffic lights.

\subsubsection{Intersection Analysis}

This module aims to gather data from multiple processed tiles for the intersections extracted in the previous section. They are categorized and then transformed into generalized intersection types. 

\begin{figure}[t] 		
	\centering
	\includegraphics{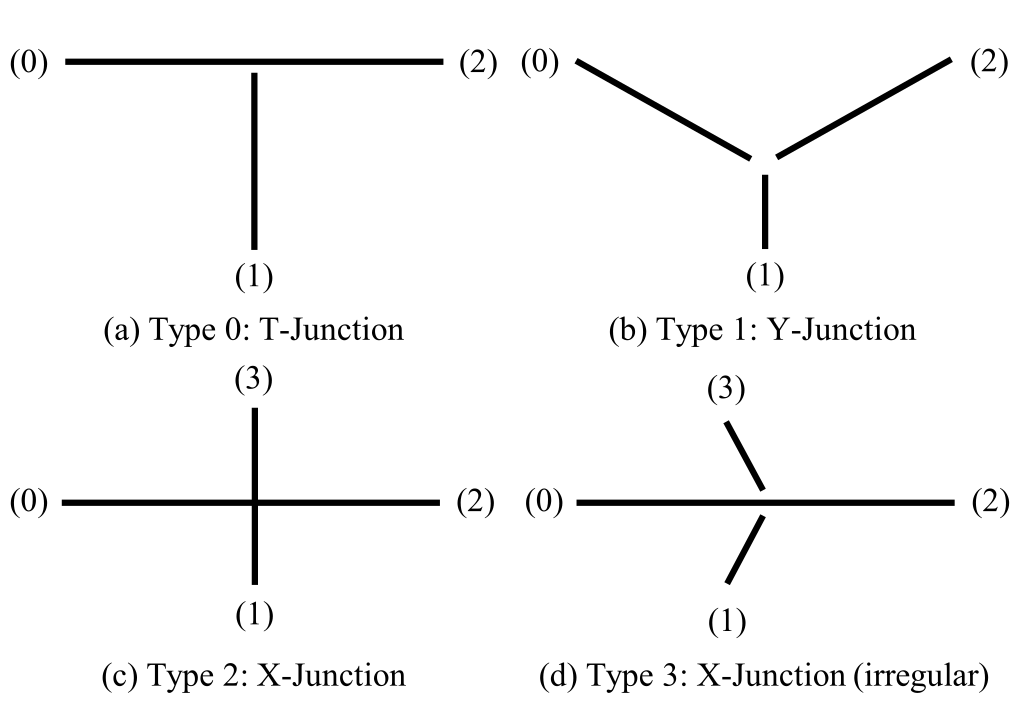}
	\caption{Intersection types are based on intersection arms orientation (numbered counterclockwise starting from the left-hand side). If two links form an angle between $150$ and $210$ degrees, they are merged into one single road (indicated by an unbroken line). The result is four different types we extract from Here HD map.}
	\label{fig_junction_types}
\end{figure}

For simplicity, only intersections with three and four arms are considered in this work. That means highway exits or roundabouts are neglected at this stage. Consequently, the remaining intersections are classified according to the types shown in Fig.~\ref{fig_junction_types}. The goal is to identify whether two opposite arms of an intersection geometrically form one continuous road, which is defined by an angle range which we set from $150$ to $210$ degree between two arms. 

Then, the generalization process follows. The goal is to rearrange the links in a way that is always identical for all intersections of a given type. For example, in a T-junction, the arm shall be index $1$ and the links forming the main road are denoted with index $0$ and $2$, respectively. This way, the statistical analysis is reproducible.

After the topology of the intersections is extracted, the lanes of the links and the lanes within the intersection are analyzed. Besides parameters such as lane width and the number of lanes, the maximum and mean curvature of turning lanes is calculated since this measure is not stored in Here HD map. The curvature is computed with the help of (\ref{eq_curv}) by calculating discrete derivatives for the centerline of each lane which is stored as a polyline.
\begin{equation}
	\kappa = \frac{x' y'' - y' x''}{(x'^2 + y'^2)^{3/2}}
	\label{eq_curv}
\end{equation}

Finally, the structured intersections are aggregated in a common data format. Table~\ref{tab_parameters} summarizes all parameters that can be extracted. Most of them are either associated with a lane on the intersection (e.g. \verb|lane_max_curvature| or \verb|lane_mean_width|) or with a link connected to the intersection (e.g. \verb|link_length| or \verb|intersection_span|). However, some parameters describe metadata of an intersection, such as the presence of traffic signals, the speed limit or its type. Those metadata can be extended if required in the future, e.g. by the distance to neighboring intersections to generate realistic road networks.

\begin{table*}[ht]
	\caption{Extracted intersection parameters from Here HD map}
	\label{tab_parameters}
	\begin{center}
		\begin{tabular}{llll}
			\textbf{Parameter} & \textbf{Scope} & \textbf{Description}\bstrut\\
			\hline
			\verb|type| & intersection & The type of the intersection denoted by an index (cf. Fig.~\ref{fig_junction_types}).\\
			\verb|traffic_signals| & intersection & Presence of traffic lights at the intersection.\\
			\verb|intersecting_angle| & link/intersection & Angle between a link and the next one, turning counterclockwise around the intersection center.\\
			\verb|intersection_span| & link/intersection & Distance from the intersection center to the start of the turning lanes coming form a link.\\
			\verb|link_length| & link & Length of a link measured from the intersection center.\\
			\verb|link_geometry| & link & Polyline of a link.\\
			\verb|Speed_limit| & link & Speed limit along a link.\\
			\verb|incoming_lanes| & link & Number of lanes corresponding to a link, approaching or leaving the intersection.\\
			\verb|lane_count| & link/intersection & Number of lanes within the intersection coming from a link.\\
			\verb|lane_mean_width| & lane & Width value of a lane within the intersection (two more variables for \verb|start| and \verb|end|).\\
			\verb|lane_max_curvature| & lane & Maximum curvature of a lane within the intersection.\\
			\verb|lane_mean_curvature| & lane & Mean curvature of a lane within the intersection.\\
			\verb|lane_heading| & lane & Heading of a lane : -1 for left, 0 for straight and 1 for right in driving direction.
		\end{tabular}
	\end{center}
\end{table*}

\section{Evaluation and Examples}\label{sec_eval}
We have applied the methodology to actual example data and evaluated the results with regard to the possible use for the road generation process. We have chosen Berlin as an example because we expect a great number of intersections concentrated in a coherent area. $49$ tiles at an area of $183.75$ square kilometers have been processed. This results in $1379$ X-junctions and $5131$ T-junctions that are extracted. The remaining two types, mentioned in Fig.~\ref{fig_junction_types}, only occur several times and are neglected for the following evaluation. 

\begin{figure}[t] 		
	\centering
    \input{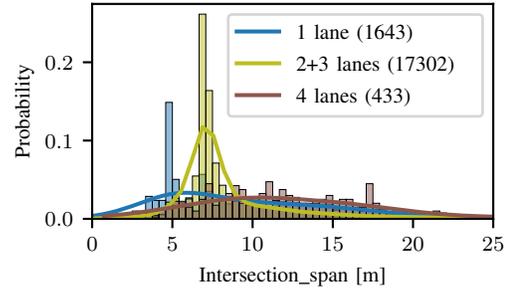}
	\caption{Intersection span of all analyzed junction arms in Berlin split according to number of lanes connected to the intersection. We grouped two and three lanes since their distribution is similar. Count of junction arms is denoted in parentheses.}
	\label{fig_interArea}
\end{figure}

Fig.~\ref{fig_interArea} shows the histogram of the intersection span of all processed junction arms. It is noticeable that the bins almost follow a normal distribution which can be exploited in the road generation process. Also, the peak at around $7$ meter for two and three lanes per junction arm is plausible for inner-city junctions. Further, it is notable that the majority of junction arms consist of two or three lanes (in and out going). Another finding is that the intersection size seems to grow with increasing number of connected lanes.

\begin{figure}[thb] 		
	\centering
    \input{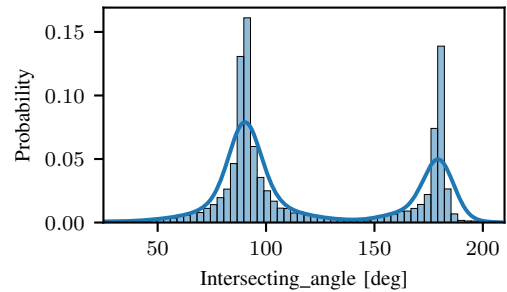}
    \caption{Intersecting angle of all analyzed T-junctions in Berlin. The angle is measured from one link to another, respectively. $n=5131$ T-junctions.}
	\label{fig_interAngleT}
\end{figure}

\begin{figure}[th] 
	\centering
    \input{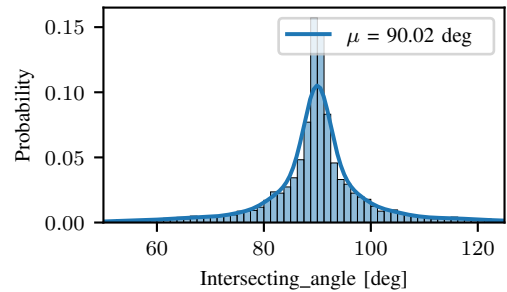}
    \caption{Intersecting angle of all analyzed T-junctions in Berlin. The angle is measured from one link to another, respectively. $n=1379$ T-junctions.}
	\label{fig_interAngleX}
\end{figure}

Another relevant parameter for an intersection design is the angle under which the junction arms intersect. Fig.~\ref{fig_interAngleT} and Fig.~\ref{fig_interAngleX} show histograms of this value for T- and X-junctions, respectively. T-junctions have two peaks at around $90$ and $180$ degrees, respectively, which matches the expectation of this type of junction. The results for X-junctions are similar, and the mean value of extracted intersecting angles is $90.02$ degree (cf. Fig.~\ref{fig_interAngleX}).


\begin{figure}[tb] 		
	\centering
    \input{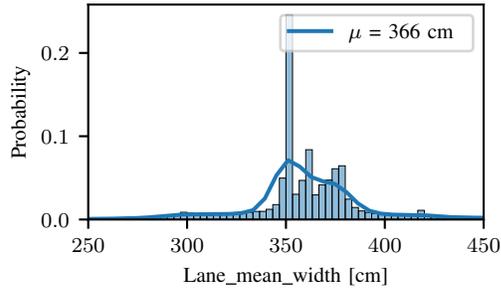}
	\caption{Mean lane width of all analyzed junctions in Berlin. $n=44265$ analyzed lanes.}
	\label{fig_interLaneWidth}
\end{figure}

A further example is plotted in Fig.~\ref{fig_interLaneWidth} where the mean width of lanes through the junction is also shown as a histogram. There is a noteworthy peak at $3.5$ meter that aligns with the road construction regulations \cite{baier2007richtlinien} for the lane width. However, the frequent occurrence of this value could also be the result of non-measured lanes, which are stored with a standard value. The mean value of $3.66$ meter seems appropriate for a metropolis such as Berlin because it is slightly wider than the $3.5$ meter from regulations. Especially when considering that the analysis of junctions in the smaller city of Aachen resulted in a mean value of $3.54$ meter for the mean lane width inside a junction. 

\begin{figure}[tb] 		
	\centering
    \input{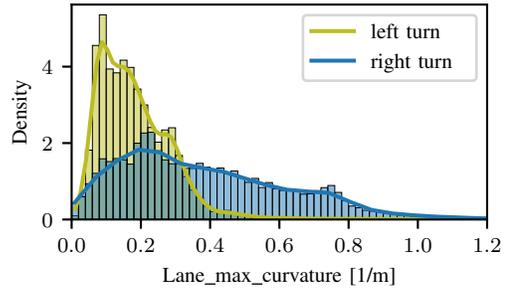}
	\caption{The density of max. curvature for turning lanes on T- and X-Junctions in Berlin. $n=13051$ and $n=14900$ for left and right turns, respectively.}
	\label{fig_interCurv}
\end{figure}

Another measure for lanes within a junction is shown in Fig.~\ref{fig_interCurv} where the maximum curvature of turning lanes is displayed. In this case, the density of the occurring values is represented to make the values comparable since there are more right-turning lanes than left turns. As expected, the maximum curvature of left turns tends to be smaller compared to right turns since their path is usually wider in countries with right-hand side traffic. The wider range of right turn curvatures may be explainable due to possibly more parallel right turn lanes. However, curvatures above $0.5 \frac{1}{\text{m}}$ seem unrealistic, which might be due to the fact that the approximation of gradients from discrete data is not that exact when there are too few points in narrow curves.

The following section describes how those results might be used to transfer them into traffic simulations.



\section{Road Network Generation and Variation}\label{sec_roadgen}
To apply the extracted road network parameters and generate a set of randomly seeded simulation maps, the existing road generator (cf.~\cite{becker2020roadgen}) has to be extended. Instead of taking only fixed values, it should be possible to use random variables when desired.

Therefore, a python wrapper has been developed that embeds the road generation tool as a C++ library and generates the required input format from a logical template. The latter has been extended so that the user can define variables used for road generation. At this point, the variables stated in Table~\ref{tab_variables} can be defined and used. In addition to Gauss and normal distribution, it is possible to define linear dependencies among variables. This allows to fulfill requirements that shall be valid no matter what the random seed of a distinct variable is. For example, if a road network of more than one junction should be generated and the angular change of the connection road between them should be fixed to $dp=\pi / 2$ radians, a simple linear equation which depends on the curvature and length of the road can be formulated:
\begin{equation}
	dp = \frac{\kappa_{s_1}}{2} l_{s_1} + \kappa_a l_a + \frac{\kappa_{s_2}}{2} l_{s_2} \text{.}
	\label{eq_90degCurve}
\end{equation}

Where in (\ref{eq_90degCurve}), $\kappa_i$ denotes curvature values and $l_i$ represents the length of a segment. The subscripts $a$ and $s_i$ label arcs and spirals, respectively. In the input template of the road variation tool (\ref{eq_90degCurve}) can be stated in straightforward mathematical notion:
\verb|dp="kS1/2 * lS1+kA * lA+kS2 * lS2"|.

\begin{table}[tb]
	\caption{Possible variable definitions for the road generation}
	\label{tab_variables}
	\def\arraystretch{1.5}
	\begin{center}
		\begin{tabular}{p{0.11\columnwidth}p{0.75\columnwidth}}
			Variable & Description \\
			\hline
			\textbf{normal} & Gauss distribution with mean value $mu$ and standard deviation $sd$. \\
			\textbf{uniform} & Uniform distribution with a min $min$ and $max$ value. \\
			\textbf{lindep} & Defines a variable that is linear dependent on other variables or constants with a dependency $dp$ in string form.
		\end{tabular}
	\end{center}
\end{table}

The resulting tool is a command-line program that takes the new input format and the desired number of outputs $n$ as inputs and creates $n$ OpenDRIVE maps as an output. During the course of the work, further compliant checks regarding the OpenDRIVE standard have been realized such as the verification of the simulation maps against the XSD schema\footnote{https://www.w3.org/XML/Schema} of OpenDRIVE. The tool allows to generate simulation maps that can be read by various simulation tools that support the OpenDRIVE standard. E.g., we imported maps into IPG CarMaker\footnote{https://ipg-automotive.com/}, RoadRunner\footnote{https://de.mathworks.com/products/roadrunner.html} and openPASS\footnote{https://openpass.eclipse.org/}.

Road networks with the four example measures and the linear dependency described in (\ref{eq_90degCurve}) have been generated and then tested in IPG CarMaker. The goal was to run a set of logical scenarios several times and examine if a prototypical driving function runs into challenging situations when the road network is varied.

\section{Conclusion \& Outlook}\label{sec_conclusion}
In this paper, we showed that characteristics of inner-city junctions can be extracted from highly detailed maps such as Here HD map. For simplicity, we categorized junctions into pre-defined types and tried to gather relevant information representing such intersection types to generate synthetic road networks from those parameter distributions. One of the results was a set of $5131$ T- and $1379$ X-junctions in the city of Berlin that corresponded to the defined types and were comparable to each other. The considered parameters showed smooth distributions, and a transfer to the conceptualized logical road description format, which was introduced in former works of the authors, seems applicable. In order to do that, an existing road generation tool was extended and is now able to generate stochastically varied road networks. 

In future work, the extracted polylines of the links' topology could be further analyzed, and geometric primitives such as lines, spirals, and arcs could be fitted through the data, which are stored only as discrete points. That would help to have a realistic and continuous curvature course between the generated intersections. However, the focus of the present work was on the characteristics of the intersections only. Further, an analysis on a more macroscopic level may be done. The density of intersections and combination of intersection type and size can be set in relation to each other and can be exploited to assemble a parametrizable, realistic composition of different junctions in a road network. Another approach that should be pursued further is the simulation setup described in Section~\ref{sec_roadgen} to systematically vary logical simulation maps and to identify relevant situations for an automated driving function.

Finally, all relevant parts of the corresponding road generator~\cite{becker2020roadgen} and the variation tool are publicly available on GitHub~\footnote{Available at https://github.com/ika-rwth-aachen/RoadGeneration} to enable further research in the generation of road layouts.



\bibliographystyle{IEEEtran}
\bibliography{roadgen, ma_wang, roadVar}

\begin{thebibliography}{10}
\providecommand{\url}[1]{#1}
\csname url@samestyle\endcsname
\providecommand{\newblock}{\relax}
\providecommand{\bibinfo}[2]{#2}
\providecommand{\BIBentrySTDinterwordspacing}{\spaceskip=0pt\relax}
\providecommand{\BIBentryALTinterwordstretchfactor}{4}
\providecommand{\BIBentryALTinterwordspacing}{\spaceskip=\fontdimen2\font plus
\BIBentryALTinterwordstretchfactor\fontdimen3\font minus
  \fontdimen4\font\relax}
\providecommand{\BIBforeignlanguage}[2]{{%
\expandafter\ifx\csname l@#1\endcsname\relax
\typeout{** WARNING: IEEEtran.bst: No hyphenation pattern has been}%
\typeout{** loaded for the language `#1'. Using the pattern for}%
\typeout{** the default language instead.}%
\else
\language=\csname l@#1\endcsname
\fi
#2}}
\providecommand{\BIBdecl}{\relax}
\BIBdecl

\bibitem{stellet2019validation}
J.~E. Stellet, M.~Woehrle, T.~Brade, A.~Poddey, and W.~Branz, ``Validation of
  automated driving--a structured analysis and survey of approaches,'' 2019.

\bibitem{wachenfeld2016release}
W.~Wachenfeld and H.~Winner, ``The release of autonomous vehicles,'' in
  \emph{Autonomous driving}.\hskip 1em plus 0.5em minus 0.4em\relax Springer,
  2016, pp. 425--449.

\bibitem{schuldt2017beitrag}
F.~Schuldt, ``Ein beitrag f{\"u}r den methodischen test von automatisierten
  fahrfunktionen mit hilfe von virtuellen umgebungen,'' Ph.D. dissertation,
  Technische Universität Braunschweig, 2017.

\bibitem{setlevel}
\BIBentryALTinterwordspacing
``Set level – simulation-based development and testing of automated
  driving,'' accessed: 2022-06-23. [Online]. Available:
  \url{https://setlevel.de/en}
\BIBentrySTDinterwordspacing

\bibitem{vvm}
\BIBentryALTinterwordspacing
``Vvm - verification and validation methods for automated vehicles in urban
  environments,'' accessed: 2022-06-23. [Online]. Available:
  \url{https://www.vvm-projekt.de/en}
\BIBentrySTDinterwordspacing

\bibitem{HanduchFASsim}
S.~Hakuli, M.~Krug, H.-P. Schöner, B.~Morys, G.~Berg, and B.~Färber,
  \emph{Handbuch Fahrerassistenzsysteme}, 3rd~ed.\hskip 1em plus 0.5em minus
  0.4em\relax Springer Verlag, 2015, ch.~II, pp. 123--164.

\bibitem{bock2018data}
J.~Bock, R.~Krajewski, L.~Eckstein, J.~Klimke, J.~Sauerbier, and A.~Zlocki,
  ``Data basis for scenario-based validation of had on highways,'' in
  \emph{27th Aachen Colloquium Automobile and Engine Technology}, 2018.

\bibitem{scholtes20206}
M.~Scholtes, L.~Westhofen, L.~R. Turner, K.~Lotto, M.~Schuldes, H.~Weber,
  N.~Wagener, C.~Neurohr, M.~Bollmann, F.~K{\"o}rtke \emph{et~al.}, ``6-layer
  model for a structured description and categorization of urban traffic and
  environment,'' \emph{arXiv preprint arXiv:2012.06319}, 2020.

\bibitem{becker2020roadgen}
D.~{Becker}, F.~{Ruß}, C.~{Geller}, and L.~{Eckstein}, ``Generation of complex
  road networks using a simplified logical description for the validation of
  automated vehicles,'' in \emph{2020 IEEE 23rd International Conference on
  Intelligent Transportation Systems (ITSC)}, 2020, pp. 1--7.

\bibitem{Opendrive}
\BIBentryALTinterwordspacing
``Asam opendrive version: 1.7.0,,'' 2021. [Online]. Available:
  \url{https://www.asam.net/index.php?eID=dumpFile
  &t=f&f=4422&token=e590561f3c39aa2260e5442e29e93f6693d1cccd}
\BIBentrySTDinterwordspacing

\bibitem{osc10}
\BIBentryALTinterwordspacing
``Asam openscenario 1.0.0 model-documentation,'' accessed: 2022-06-23.
  [Online]. Available:
  \url{https://releases.asam.net/OpenSCENARIO/1.0.0/Model-Documentation/content/OpenScenario.html}
\BIBentrySTDinterwordspacing

\bibitem{asamOpenX}
B.~Engel and N.~Dillmann, ``Asam openx,'' Online Presentation, 10 2020.

\bibitem{hillerRoadAggr}
J.~Hiller, F.~Müller, and L.~Eckstein, ``Aggregation of road characteristics
  from online maps and evaluation of datasets,'' in \emph{2021 IEEE Intelligent
  Vehicles Symposium (IV)}, 2021, pp. 208--214.

\bibitem{Despine11}
G.~Despine and C.~Baillard, ``Realistic road modelling for driving simulators
  using gis data,'' pp. 431--448, 01 2011.

\bibitem{Cura15}
R.~Cura, J.~Perret, and N.~Paparoditis, ``Streetgen: In-base procedural-based
  road generation,'' \emph{ISPRS Annals of Photogrammetry, Remote Sensing and
  Spatial Information Sciences}, vol. II-3/W5, pp. 409--416, 08 2015.

\bibitem{Li20}
\BIBentryALTinterwordspacing
Q.~Li, Z.~Peng, Q.~Zhang, C.~Qiu, C.~Liu, and B.~Zhou, ``Improving the
  generalization of end-to-end driving through procedural generation,''
  \emph{CoRR}, vol. abs/2012.13681, 2020. [Online]. Available:
  \url{https://arxiv.org/abs/2012.13681}
\BIBentrySTDinterwordspacing

\bibitem{Pyodrx}
\BIBentryALTinterwordspacing
``Pyodrx,'' accessed: 2022-06-23. [Online]. Available:
  \url{https://github.com/pyoscx/pyodrx}
\BIBentrySTDinterwordspacing

\bibitem{muktadirrealistic}
G.~M. Muktadir, A.~Jawad, A.~Shepelev, I.~Paranjape, and J.~Whitehead,
  ``Unpublished: Realistic road generation: Intersections,'' 5 2022.

\bibitem{baier2007richtlinien}
R.~Baier and A.~S.~F. f{\"u}r Stra{\ss}en-und, \emph{Richtlinien f{\"u}r die
  Anlage von Stadtstra{\ss}en: RASt 06}.\hskip 1em plus 0.5em minus 0.4em\relax
  FGSV Verlag, 2007.

\bibitem{heremaps}
{HERE Global B.V.}, ``Here hd live map (onepager),'' 2020.

\end{thebibliography}

\end{document}